# LBCIM: Loyalty Based Competitive Influence Maximization with $\epsilon$-greedy MCTS strategy

Malihe Alavi*, Farnoush Manavi, Amirhossein Ansari, Ali Hamzeh

**Abstract**
Competitive influence maximization has been studied for several years, and various frameworks have been proposed to model different aspects of information diffusion under the competitive environment. This work presents a new gameboard for two competing parties with some new features representing loyalty in social networks and reflecting the attitude of not completely being loyal to a party when the opponent offers better suggestions. This behavior can be observed in most political occasions where each party tries to attract people by making better suggestions than the opponent and even seeks to impress the fans of the opposition party to change their minds. In order to identify the best move in each step of the game framework, an improved Monte Carlo tree search is developed, which uses some predefined heuristics to apply them on the simulation step of the algorithm and takes advantage of them to search among child nodes of the current state and pick the best one using an $\epsilon$-greedy way instead of choosing them at random. Experimental results on synthetic and real datasets indicate the outperforming of the proposed strategy against some well-known and benchmark strategies like general MCTS, minimax algorithm with alpha-beta pruning, random nodes, nodes with maximum threshold and nodes with minimum threshold.

## 1 Introduction

Social networks have been used to analyze societies' structures. The studies on these networks are, in fact, the study of the desired society's interpersonal relationships and an investigation on how these networks are constructed (Borgatti et al. 2018). Influence maximization (IM) is one of the most helpful understudy fields in social networks, first proposed by Kempe et al. (2003). The main goal that IM tries to achieve is summarized in finding the subset of k node (seed set), which cause the most influence spread through the network, that means the number of the infected nodes in the desired network maximized. Viral marketing or information spread by word-of-mouth is one of the essential applications in IM (Domingos et al. 2001). A study declares that people consult with their friends and family before buying a new product, especially when it comes to digital products (Burke et al. 2003). Some other applications of IM can be pointed as network monitoring (Leskovec et al.2007), social recommendation (Ye et al. 2012), and rumor control (Budak et. al. 2011). Many related IM studies use the two crucial diffusion models: the linear threshold model (Granovetter et al. 1978) and the independent cascade model (Goldenberg et al.2001).

The studies in IM assume that one party tries to maximize its influence through the network, and no rival party exists, while in reality, two or more parties compete. They try to maximize their influence or minimize the opponent's influence (Brede et al. 2018) or even minimize the cost of influence spread by focusing on the nodes that affect seed nodes (Talukder et al. 2019). Explicit examples of this situation are political elections or competition between companies for selling their products like Apple and Samsung. In such competitive environments, each party, in addition to maximizing its influence, tries to minimize the rival influence through the network; therefore, Competitive Influence Maximization (CIM) is defined as finding the best strategy that has the best performance considering the rivals' strategies.

Generally, there are two scenarios in CIM that causes different seed selection strategies (Ansari et al. 2019). In the first scenario, parties choose their seeds regardless of the opponent's turn (Ansari et al. 2019, Bhowmick et al. 2015, Ali et al. 2018), and the second scenario follows turn-based algorithms that each party waits for the opponent to make his selection and then takes his shot (Bozorgi et al. 2017, Li et al. 2018, Yan et al. 2019).

The most critical issues CIM has dealt with in recent years can be pointed to the problems that incorporate time factor to their model and consider a time limitation on influence spread (Pham et al. 2019, Ali et al. 2018, Li et al.2018), the issues that investigate the role of trust/ distrust between users in the CIM model (Caliò et al. 2019, Zhang et al. 2019, Wang et al. 2021, Asadi et al. 2021). Some other studies aim to find a solution for CIM in online social networks (OSN); these problems arise when there is not enough network structure information, therefore one tries to find the unknown parameters of network through the learning strategies or propose a reasonable solution based on these incomplete pieces of knowledge of network. (Zuo et al.2020, Pham et al. 2018, Pham et al. 2019, Pham et al 2019)

M. Alavi is the corresponding author.
M. Alavi, F. Manavi and A. Hamzeh are with the Shiraz University, Department of Computer Science and Engineering, Shiraz, Fars, Iran (e-mail: malihealavi@cse.shirazu.ac.ir, f.manavi@cse.shirazu.ac.ir, ali@cse.shirazu.ac.ir ).
A. Ansari was with the Shiraz University Department of Computer Science and Engineering, Fars, Iran. He is now with the Simon Fraser University, British Colombia, Canada (e-mail: a.ansari@cse.shirazu.ac.ir)

One of the outstanding shortcomings in previous studies can be pointed out as the lack of nodes' loyalty in the CIM environment. Whereas, nodes' loyalty is essential in determining a party's victory or defeat. Just in (Ishay et al. 2018), two CIM variants have been proposed which are based on loyalty; one of them considers zero-loyalty attribute for all nodes of the network, so nodes turn to inactive state right after activation without even the opponent trying to change the node's state; therefore, nodes can participate in the game as long as the game is continued. The second variant considers nodes to be fully loyal and causes a node to remain activated after activation; therefore, the activated nodes are excluded from the competition to the end of the game. Both of these situations are unrealistic and far from the truth.

Being utterly loyal to a party or a product seems unlikely to happen. Based on (Gu et al. 2009), factors like customer satisfaction, perceived utilitarianism, hedonic value, and level of customer knowledge about alternative products and services determine a party's loyalty level. Whether the opponent proposes better suggestions or changes the environment factor, people may change their thoughts about their favorite brand or party and follow the opponents. The history of Nokia mobiles is a clear example of the claim mentioned above. When Nokia mobiles entered the markets, they were very well received. So many mobile users preferred to use Nokia mobiles, and they took the majority of mobile marketing. However, with the smart-phones developments equipped with new capabilities and lack of precise technology forecasting by Nokia, Nokia fans were reduced ,and smartphone fans were added (Alibage et al. 2018). Another example is the presidential election of the USA in the years 2016 and 2020, wherein 2016 Donald Trump as a candidate of the republican party, succeeded to win by 306 electoral votes to 232 electoral votes for his opponent Hilary Clinton who was a candidate of the democratic party. Despite most predictions considering Clinton as the ultimate winner of the election, but Trump won the competition in 2016, while in 2020, with the appearance of COVID-19, the US government needs to change its politics to overcome this issue; besides, Biden as a presidential competitor, proposed some suggestions for dealing with the corona concern, which was the main reason causing Trump to lose the competition in 2020. Other reasons like economic distress, demographic insensitivity, poor event management, and overweening narcissism have an influential role in the election result in 2020 (Hart et al. 2022). All these environmental factors caused in some USA states like Arizona, Michigan, Pennsylvania, Georgia, and Wisconsin people change their minds. Despite their votes in 2016, which were in Trump's favor, their electoral votes were in Biden’s favor in 2020. (2020)

It is worth mentioning that people, usually affected by a party or a brand, may change their minds harder than before because they feel some sort of fixation with the previous party (Ruiz-Mafe et al. 2014). Moreover, relatives and friends also considerably impact one's changing opinion. The more someone's relatives are fans of one party, the more difficult it is to change his mind in favor of another party.

A loyalty-based framework is proposed that considers a competitive turn-based game between the two parties, red and black, to tackle the earlier challenges. The main contribution of this paper is as follows:

1) Proposing Loyalty Based Competitive Influence Maximization (LBCIM), a CIM framework for competition between two parties, A and B, over the members who are not entirely loyal to their selected party, and if the opponent party proposes better suggestions to them, they might change their mind.

2) The competition is turn-based, so each time it comes to a player's turn, he can choose one node among the $\{inactive \cup active\ for\ opponent\}$ nodes.
3) A diffusion model is proposed based on a linear threshold model but is extended to be appropriate for the LBCIM framework. In this model, a threshold is assigned to each node equal to nodes' degree; whenever a node becomes activated for a party, its threshold increases proportionally to the number of node's neighbors.
4) A set of appropriate and consistent heuristics are defined and used in Monte Carlo Tree Search (MCTS) algorithms as a strategy for finding the best node. Reported results in the results section demonstrate the outperform of this strategy.

The rest of the paper is organized as follows: section two presents the related works. In section three, backgrounds of our study are presented. In section four, the proposed framework and strategy are described in detail. In section five, empirical results are reported, and in the last section, the conclusion and future works are discussed.

## 2 Related Works

In this section, related works of recent years are reviewed. These studies are divided into two parts; in the first part, CIM related works are studied and in the second part, the related works of MCTS are reviewed.

### 2.1. Competitive Influence Maximization

In (Ishay et al. 2018), a CIM zero-sum game with two player competition is proposed. Both of the players have the equal amounts of tokens, and in each game round, they choose a node to put a token on. When the number of tokens of a node reaches to the node threshold, the node will activate and then it spreads all of its tokens to its neighbors. Alpha-Beta pruning and Monte Carlo tree search algorithms have been used for finding the optimal strategy for the players.

In (Pham et al. 2019), instead of maximizing the profit, a model for minimizing the cost of CIM problem under the dynamic network has been proposed. The authors focused on the nodes that directly effect on the seed nodes, and tried to find such these nodes to minimize the cost of seed activation. In (Xie et al. 2021), the authors claimed that the active nodes only effect on the local community around them not the whole network so they proposed a two-phase seed selection to select the minimum number of nodes with maximum range of influence to cover all the communities. They also defined a propagation model that in addition to considering the active nodes influence propagation, it designates the inactive nodes for propagating their influence.

In (Niknami et al. 2022), a competitive framework has been proposed that in addition to seed selection, budget allocation is also considered and these two phases are integrated and are used in the Reinforcement Learning model. In each round of the game, the influence weights of all nodes are calculated based on the most reliable influence path (MRIP) of each node and its weight. MRIP declares the shortest path between two nodes with the highest edges probabilities. The node with the highest influence weight is chosen as a new seen node. For the budget allocation phase using an RL agent, one can decide to allocate budget to new seed node or feed the previous activated node in order to increase its influence on neighbors.

### 2.2 Monte Carlo tree search

There have proposed so many extensions on the original MCTS and they've tried to improve the functionality and precision of the algorithm. In (Baier et al. 2014), the authors presented three extensions for MCTS for employing heuristic knowledge within MCTS. Their approaches are called MCTS with informed rollouts (MCTS-IR), MCTS with informed cutoffs (MCTS-IC) and MCTS with informed priors (MCTS-IP), where checks out the role of the tree height limitation, MCTS simulation policy and MCTS and minimax hybrids. In (Lanctot et al. 2014) a new technique named implicit minimax backups has been introduced to store the information from both MCTS and minimax separately and combining these two separate sources to define a new selection policy. In fact, with an implicitly-computed minimax search which uses heuristic evaluations, they augment the quality of MCTS simulations. these heuristic evaluations are used as separate source of information, and backed up in the same way as in classic minimax search

## 3 Background

In this section, the prerequisites concepts of LBCIM framework are described.

### 3.1 General MCTS

The original MCTS algorithm consists of 4 steps executed iteratively until termination criteria are realized. These steps are described briefly in the following (Baier et al. 2014):

1) Selection: in this step, a predefined selection policy traverses the tree and the most urgent node is selected. The selection policy evaluates the tree nodes based on an evaluation function and prioritizes them based on their estimated values. The most common selection policy is *UCB1 multi-armed bandit*, which is also used in this paper.

$$UCT = \overline{x_i} + c.\sqrt{\frac{\ln(n_v)}{n_i}} \quad (1)$$

Where in (1), $\overline{x_i}$ is the node's average score (winning rate), $n_i$ is the number of times that node $i$ has been visited, $n_v$ is the number of times that the parent (node $v$) has been visited, and $c$ is a constant for tuning the amount of exploration and exploitation.

2) Expansion: after selecting a node in the selection phase, one or more children of the selected node are added to the tree. In this paper, one node is added in this step.
3) Simulation: in this step, using a predefined rollout policy, the game is simulated from the newly added node to the end of the game. Rollout policy can be weak, like choosing nodes randomly or be strong, like using strategies and knowledge while simulating the game. A strong policy helps reduce the execution time and find more beneficial nodes.
4) Backpropagation: based on the results of the simulation step, the value of newly added node is determined, then the values of the other nodes should be updated beginning from the new node to the root of the tree.

## 4 Proposed Method

In this section, proposed Loyalty Based Competitive Influence Maximization (LBCIM) framework and the proposed strategy are explained in detail. In the first subsection, the game framework and its dependencies, including the game board, diffusion model, loyalty factor, game process and the termination criteria, are described. In the next subsection, the concept of $\epsilon$-greedy MCTS and the working heuristics are examined.

### 4.1 The LBCIM Framework

#### 4.1.1 Game Board

The intended network is implemented by a simple undirected graph $G = (V, E)$, where $V$ represents the nodes and $E$ represents the graph's edges. Each node $u$ contains three values $(\theta_u, T_u^r, T_u^b)$, where these values represent the threshold of node $u$, number of red tokens, and number of black tokens that node $u$ has earned, respectively. The edges of the graph are undirected and weightless, and for more simplicity, competition of just two parties has been studied. Each node can be in three states: $S = \{inactive, Red, Black\}$. According to (2) and (3), at the beginning of the game, all nodes are inactive, and the threshold of each node is equal to its degree.

$$S_v = inactive\ for\ all\ v \in V \quad (2)$$

$$\theta_v = D_v\ \ for\ all\ v \in V \quad (3)$$

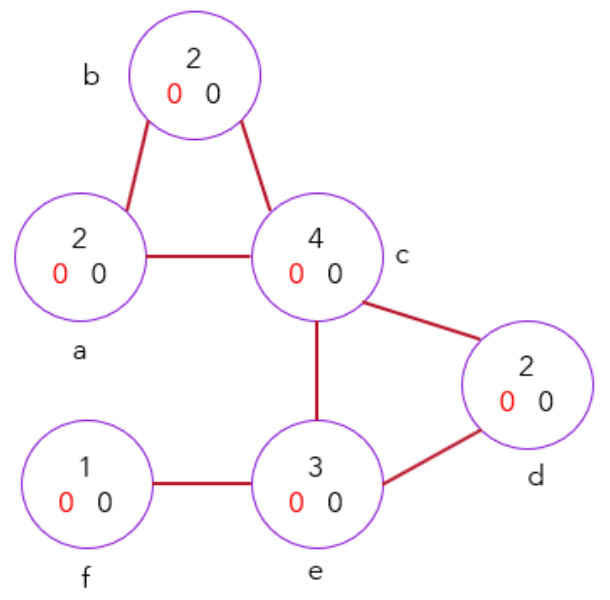

**Fig. 1.** A sample graph with six nodes representing the gameboard

Fig. 1, demonstrates a simple gameboard with six nodes and seven edges. The value at the top of the nodes indicates $\theta$, the left indicates $T^r$, and the right indicates $T^b$. Since this paper focuses on fair competition, each party is given $\left\lfloor |v| - \frac{|E|}{2} \right\rfloor$ tokens (Ishay et al. 2018). The tokens determine the maximum number of choices each party can has. These tokens can be considered advertisements or free products that the companies give to the influencers to attract people.

### 4.1.2 Diffusion Model

LBCIM diffusion model is an extended version of the linear threshold model that considers loyalty. In the LBCIM diffusion model, all edges are assigned the same weight, and according to (4), node $u$ of the graph turns its state from inactive to $Red/Black$ when the desired circumstances come true.

$$\begin{cases} \textbf{red} \quad if\ \left(T_u^{\mathrm{r}} + T_u^{\mathrm{b}}\right) \geq \theta_u \ \textbf{and}\ \left(T_u^{\mathrm{r}} > T_u^{\mathrm{b}}\right) \\ \qquad \textbf{or}\ \left[T_u^{\mathrm{r}} = T_u^{\mathrm{b}}\ \textbf{and}\ lastToken = red\right] \\ \\ \textbf{black} \quad if\ \left(T_u^{\mathrm{r}} + T_u^{\mathrm{b}}\right) \geq \theta_u \ \textbf{and}\ \left(T_u^{\mathrm{b}} > T_u^{\mathrm{r}}\right) \\ \qquad \textbf{or}\ \left[T_u^{\mathrm{r}} = T_u^{\mathrm{b}}\ \textbf{and}\ lastToken = black\right] \end{cases} \tag{4}$$

If a node receives an equal amount of red and black tokens, it changes its state based on the color of the last token it has earned. After a node is activated, all the tokens it has received are changed to the color of the node state, e.g., $Red$ or $Black$. The activated node then diffuses all its tokens among its neighbors. Algorithm 1 describes the LBCIM diffusion model approach.

### 4.1.3 Loyalty Factor

After a node activates, its threshold increases by $\alpha$ to represent its dependency on its party and will be reactivated more difficult than before.

$$\theta_{new} = \theta_{old} + \alpha, \left(\alpha = \left\lceil \frac{num.\ of\ neighbor}{4} \right\rceil\right) \tag{5}$$

Equation 5 indicates the adjusted value for $\alpha$ in this paper. Experiments on different values of $\alpha$ demonstrate that high values of $\alpha$ resulting too many threshold's increments after each activation, incapables the opponents to reactivate the node; therefore, the node will be removed from the competition. This situation is the same as *the full-loyalty variant* of (Ishay et al. 2018). On the other hand, small alpha values resulting in a bit threshold's increments in each time, facilitating nodes' reactivation for the opponents. It is the same as the zero-loyalty variant of (Ishay et al. 2018).

### 4.1.4 Game Process

The main process of the game is turn-based. When it comes to the black (red) player, he chooses a node $u$ among the $\{inactive \cup red(black)\}$ nodes with his strategy; if he has sufficient tokens to activate the selected node, he will donate the proper number of tokens to the selected node to activate it; otherwise, he only gives one token to the node $u$, hoping that it might be activate in the following rounds.

If the node $u$ got activated after receiving tokens, one of the following conditions occurs:

a) The previous state of node $u$ was $inactive$: in this case, nodes' threshold is equal to its degree, and it has received some tokens equal to its degree. So, in the diffusion phase, it gives one black(red) token to each of its neighbors.

$$\textbf{if}\ \theta_u = D_u\ \textbf{then}: \tag{6}$$
$$(T_n^b)_{new} = (T_n^b)_{old} + 1 \ for\ n \in \{neighbors\ of\ u\}$$

b) The previous state of node $u$ was $Red$ $(Black)$: in this case, the node $u$ has a threshold more than its degree, so the tokens it has earned are also more than its neighbors. Therefore, the activated node $u$ diffuses its tokens to neighbors according to their threshold (descending order). The diffusion process between the ordered neighbors will go on until the tokens of the activated node finish.

$$\textbf{if}\ \theta_u > D_u\ \textbf{then}: \tag{7}$$
$$\begin{cases} (T_n^r)_{new} = (T_n^r)_{old} + \left\lfloor \frac{\theta_u}{D_u} \right\rfloor + 1 \quad for\ the\ |\ \theta_u\ mod\ D_u|\ neighbors\ with\ the\ highest\ threshold \\ \\ (T_n^r)_{new} = (T_n^r)_{old} + \left\lfloor \frac{\theta_u}{D_u} \right\rfloor \quad for\ the\ rest\ of\ the\ neighbors \end{cases}$$

When the diffusion process is done, the desired node's threshold is increased by $\alpha$. It is worth mentioning that each node's neighbors may be activated in the diffusion process. If multiple nodes get ready to be activated, first, they are arranged in descending order based on the difference between the black and red tokens (red and black tokens), and after that, they get activated and diffuse their tokens, respectively. The firing and diffusion process may be continued for multi-levels hierarchically. When no node is left to be fired, the game is left to the red (black) player. Algorithm 2 describes the LBCIM game framework.

### 4.1.5 Termination Condition

The game will be over when there left no more tokens in the player's hand. If one of the players has finished their tokens and the other still has some tokens, the game will be continued, and the player with tokens chooses the nodes continuously until his tokens run out.

### 4.1.6 Game Winner

When the game ends, the winner of the game will be determined based on the activated node number. The player who activates more nodes wins the game.

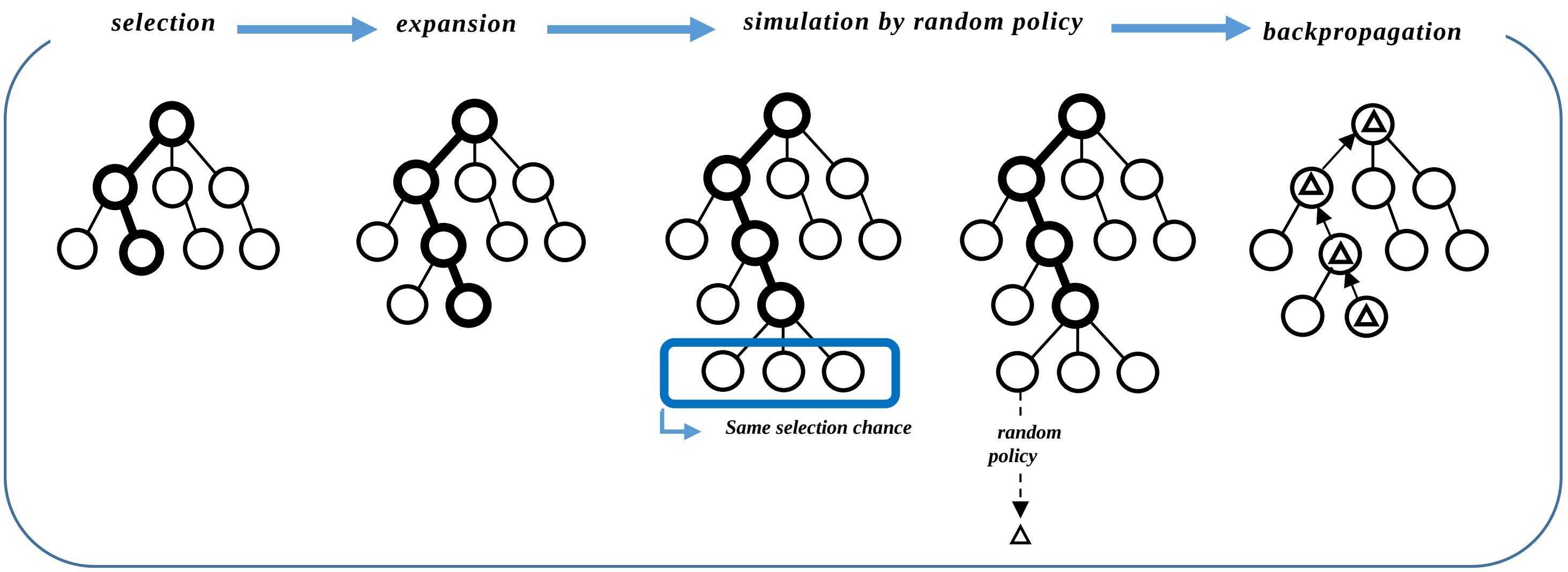

**Fig. 2.** Main steps of general MCTS strategy

**Algorithm 1:** LBCIM diffusion model

**Input**: graph, activated node;
**variables**: u = activated node, neighbors = neighbors of activated node;
**if** $S_u$ == *red*
$T_u^r \leftarrow T_u^r + T_u^b$ *// black tokens of node u are changed to red tokens*
**if** $\theta_u == D_u$ **then**
**for** n in *neighbors* **do**
$T_n^r \leftarrow T_n^r + 1$
$T_u^r \leftarrow T_u^r - 1$
**end for**
**else if** $\theta_u > D_u$ **then**
*sorted_neighbors* ← *sort the neighbors based on θ in descending order*
**repeat:**
**for** n in *sorted_neighbors* **do**
$T_n^r \leftarrow T_n^r + 1$
$T_u^r \leftarrow T_u^r - 1$
**if** $T_u^r == 0$ **then** break
**end if**
**end for**
**until** $T_u^r > 0$
**end if**
**end if**

**Algorithm 2:** loyalty based CIM game framework

**Input**: a simple undirected graph, player
**variables**: *selected_node* = *null* , *node_capacity*
**if** *player* == *red*
**if** *player_tokens* > 0 **do**
**repeat**
*selected_node* ← ***null***
**if** there remain any unselected node **do**
$selected_{node} \leftarrow$ *choose a node among {inactive ∪ black} using the selected strategy*
$node_capacity \leftarrow \theta_{selected\ node} - (T^r_{selected_node} + T^b_{selected_node})$
**else**
*break*
**end if**
**until** player tokens > *node_capacity*
**if** *selected_node* ! = *null* **do**
$T^r_{selected_node} \leftarrow T^r_{selected_node} + node_capacity$
*player token* ← *player token* − *node_capacity*

### 4.2 Game Solution Strategy: Heuristic-based ϵ-greedy MCTS

As mentioned in section 3, MCTS rollout policy has a considerable impact on the output utility of the MCTS algorithm. Accordingly, some compatible heuristics have been defined and used in the rollout policy of the algorithm. Indeed, as it can be concluded from Fig. 2 and 3, instead of choosing the simulation moves randomly, they are evaluated with these heuristics; the best of them is selected with probability $\epsilon$, or one of the others is selected with probability $1 - \epsilon$. The best value of $\epsilon$ has been obtained by test and trial on different values. Based on the experiments, $\epsilon = 0.7$ performs better than the other values. Algorithm 3 describes the heuristic based $\epsilon$-greedy MCTS approach.

The heuristics for the black player are described in the following:

1) **P**arity **H**euristic (**PH**): prioritizing the nodes with more difference between black and red tokens.

$$argmax_v\ (T_v^b - T_v^r) \tag{8}$$

2) **H**ubs with **W**eak **N**eighbor (**HWN**): prioritizing the nodes with many neighbors with a low degree. It is more likely that some of their neighbors get activated by activating these nodes.

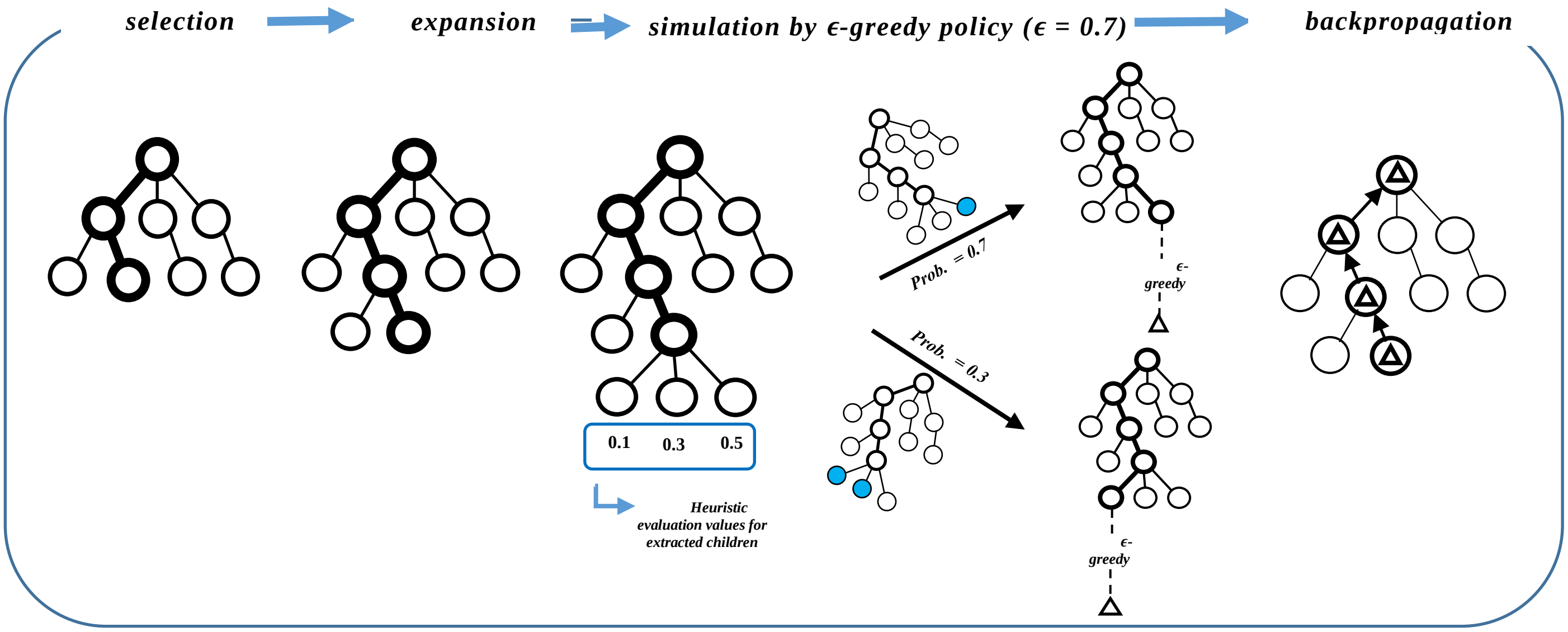

**Fig. 3.** Main steps of $\epsilon$-greedy MCTS strategy

3) **N**odes with **L**ow **T**hreshold (**NLT**): prioritizing the nodes with a low degree. In this heuristic, nodes with *Red* state are given more scores than *inactive* nodes.
4) **H**ubs on the **V**erge of **A**ctivation (**HVA**): prioritizing the nodes with a high degree (many neighbors) and low capacity for firing. These nodes need a little effort to be activated and influence many other nodes.

Eventually, a linear combination of these four heuristics with the same weights is used for evaluating the nodes in MCTS.

$$final_metric = \alpha.PH + \beta.HWN + \gamma.NLT + \lambda.HVA \quad (9)$$

$$(\alpha = \beta = \gamma = \lambda = 0.25)$$

**Algorithm 3:** heuristic based $\epsilon$ greedy MCTS

```
Input: current state of game environment as root of tree
variables: selected_node = null , node_capacity
function main(root):
    while resource_left(time, itr_num) do
        leaf ← traverse(root)
        simulation_result ← rollout(leaf)
        backpropagate (leaf, simulation_result)
    end while
    return best_child (root)
function traverse(node):
    while fully_expanded(node) do
        node ← best_uct(node)
    end while
    return pic_unvisited(node.children) or node
function rollout(node):
    while non_terminal(node) do
        node ← rollout_policy(node)
    end while
    return result(node)
function rollout_policy(node):
    evaluated_nodes ←
        heuristic_evaluation(node.children)
    chosen_node ← pick_greedy(evaluated_nodes)
    return chosen_node
function pick_greedy(nodes, ϵ):
    node ← pick the best evaluated node
        with prob ϵ or one other node randomly
        with prob 1 − ϵ (nodes)
    return node
function backpropagate(node, result):
    if is_root(node)
        return
    end if
    node.stats ← update_stats(node, result)
    backpropagation (node.parent)
function best_child (node):
    pick child with highest number of visits
```

## 5 Experimental Results

### 5.1 Dataset and System Configuration

The source code is run on a system with 32 GB of RAM and core i5 CPU and win 10 as OS. Python has been used as the programming language, and synthetic datasets have been created with the help of the *networkx* library. The source code is accessible from project repository[1].

In order to evaluate the LBCIM framework with $\epsilon-$ *greedy MCTS* strategy, three synthetic datasets and two real datasets have been used. The specification of these datasets is described in the following.

[1] https://github.com/malihee/LBCIM

### 5.1.1 Synthetic Dataset

a) Random Graphs

In order to create Random Graphs, *Erdős–Rényi* model has been used[2]. In this model, $G(n, p)$ represents a graph with $n$ nodes that the probability of edge creation between every two nodes is equal to $p$. In other words, the probability of creating a graph with $n$ nodes and $M$ edges, is as follow:

$$p^M(1-p)^{\binom{n}{2}-M} \tag{10}$$

Fig. 4 demonstrates an example of random graphs created with *Erdős–Rényi* model.

b) Scale Free Graphs

In these graphs, the graph structure is independent of graph size, which means that with the growth of the network, its structure stays the same. In this paper, the *barabasi-albert* model has been used[3]. Fig. 4 demonstrates an example of scale-free graphs created with the *barabasi-albert* model.

c) Small-world Graphs

Most nodes do not connect in these graphs, but neighbors of a node are usually neighbors themselves. In this paper, the *watts-strogatz* model has been used for creating the small-world graph[4]. Fig. 4 demonstrates an example of small-world graphs created with the *watts-strogatz* model.

**Table 1** Parameters of synthetic dataset

| | $n$ | $m$ | $k$ | $p$ |
|---|---|---|---|---|
| Small world | $rand(30,100)$ | ----- | $rand(3,7)$ | 0.3 |
| Scale free | $rand(30,100)$ | $rand\left(1, \left\lfloor \frac{n}{3} \right\rfloor\right)$ | ----- | ----- |
| Random graph | $rand(30,100)$ | ----- | ----- | 0.3 |

Table 1 provides information about the parameters that have been used in *networkx* library to generate the synthetic dataset. In the first row of Table 1, *n*, *m*, *k*, and *p* determine the number of nodes, number of edges to attach from a new node to existing nodes, number of nearest neighbors of each node in a ring topology, and probability of link creation respectively.

### 5.1.2 Real Dataset

a) Facebook

The Facebook dataset contains 4039 nodes and 88234 weightless edges[5]; these edges represent the friendship between the nodes in the Facebook network. In order to test the LBCIM framework on this dataset, one cluster of the graph with 198 nodes is selected using the label propagation method, and 100 nodes of these 198 nodes are picked to be tested. Fig. 4 demonstrates the selected sample with 100 nodes of the Facebook graph.

b) Peer to Peer

This dataset contains 62586 nodes and 147892 weightless edges[6]. Label propagation method is also performed on this dataset, and a cluster with 100 nodes is picked. Fig. 4 demonstrates the selected sample with 100 nodes of the peer-to-peer graph.

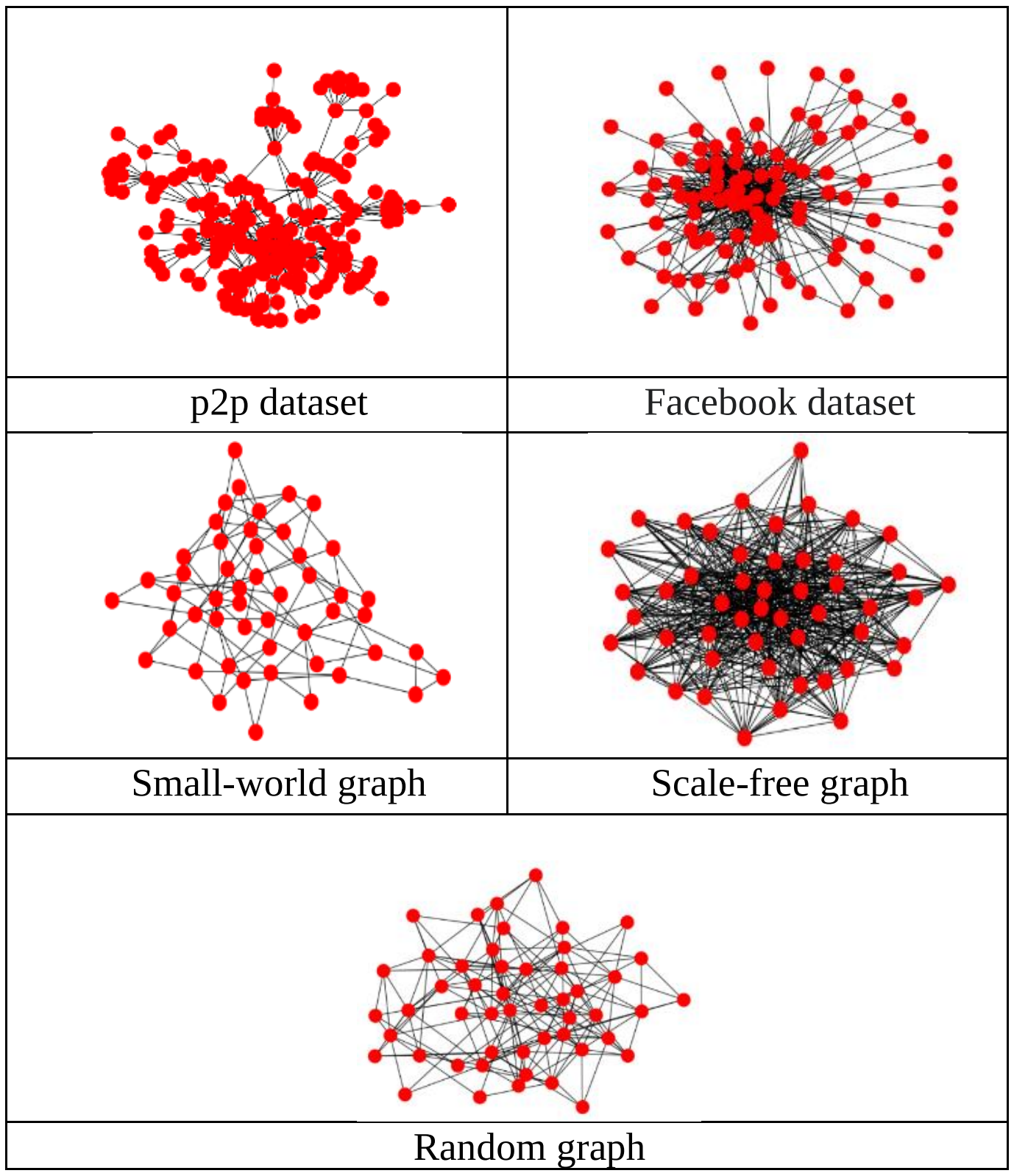

**Fig. 4:** Graph samples with 100 nodes.

### 5.2 Evaluation Metrics and Experimental Results

Common evaluation metrics have been used to evaluate the proposed method and compare it with the opponent strategies. If $w$ is the number of times that black ($\epsilon - greedy\ MCTS$) wins and $l$ is the number of times that red (opponent strategy) wins, and $d$ is the number of times that none of the players win the competition (draw), then evaluation metrics are defined as follows:

$$win\ rate = \frac{w}{w+l+d} \tag{11}$$

$$loss\ rate = \frac{l}{w+l+d} \tag{12}$$

$$draw\ rate = \frac{d}{w+l+d} \tag{13}$$

[2] https://networkx.org/documentation/networkx-1.10/reference/generated/networkx.generators.random_graphs.erdos_renyi_graph.html

[3] https://networkx.org/documentation/networkx-1.9.1/reference/generated/networkx.generators.random_graphs.barabasi_albert_graph.html

[4] https://networkx.org/documentation/networkx-1.9/reference/generated/networkx.generators.random_graphs.watts_strogatz_graph.html

[5] https://snap.stanford.edu/data/ego-Facebook.html

[6] https://snap.stanford.edu/data/p2p-Gnutella31.html

The diagrams 4-9 show the competition between the black player's strategy, which is $\epsilon - greedy\ MCTS$, and the red player's different strategies. Red player's strategies include general MCTS, Minimax algorithm with alpha-beta pruning with length 4, choosing a node at random, choosing a node with a minimum threshold, and choosing a node with the maximum threshold in each turn. The heuristics that are used in minimax algorithm are the heuristics that are defined in subsection 4.2.

Each experiment has been run 100 times; in 50 runs, black player starts the game, and other 50 runs, red player starts. Also, a new graph is created in the synthetic dataset for each run to differentiate its structure from the previous ones. The amount of randomness in the results is also reported in 2-5 that will be explained in detail in the following.

**Table 2** Randomness of Small-World dataset on five graphs

| **Randomness** | **Graph 1** | **Graph 2** | **Graph 3** | **Graph 4** | **Graph 5** |
|---|---|---|---|---|---|
| Red Wins (%) | 10 | 20 | 20 | 10 | 10 |
| Black wins (%) | 85 | 70 | 75 | 80 | 85 |
| Draw (%) | 5 | 10 | 5 | 10 | 5 |

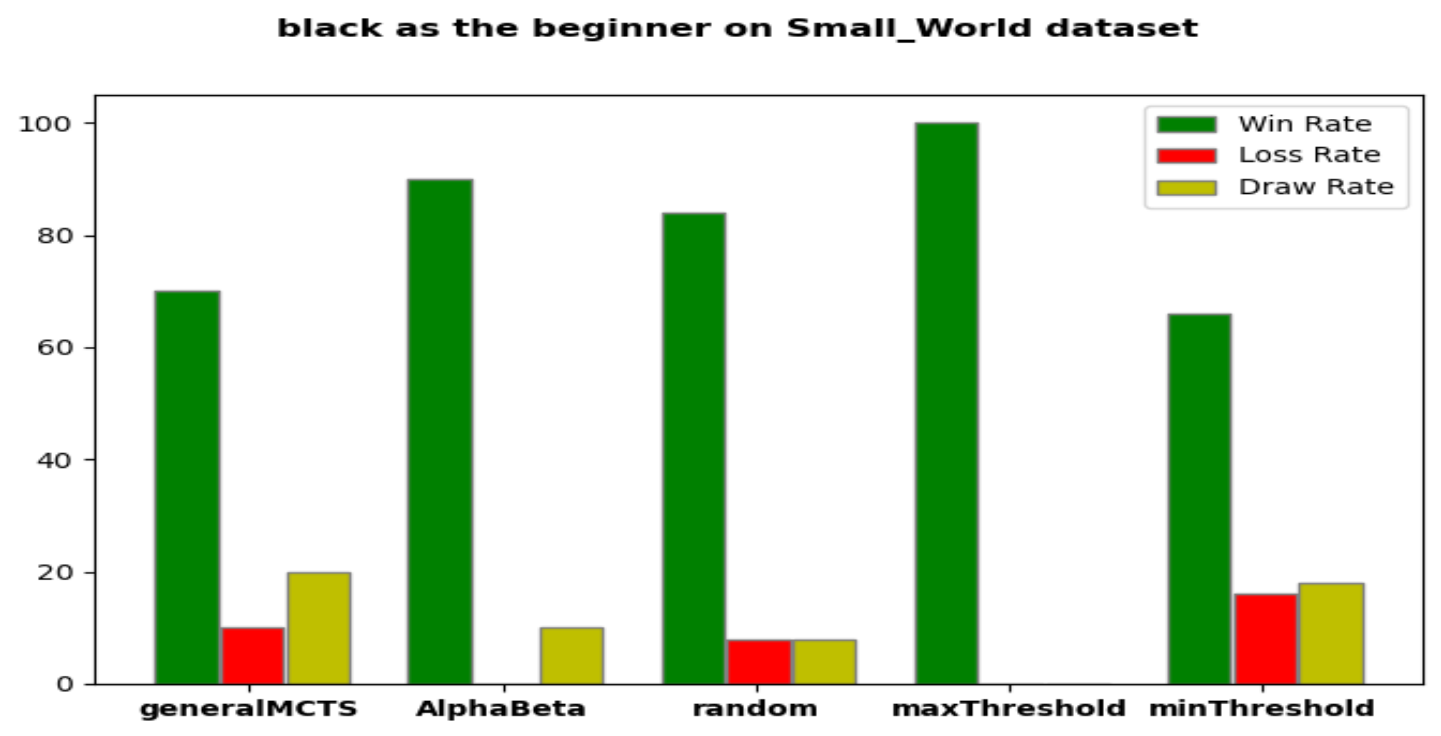

(a) Black player starts the game

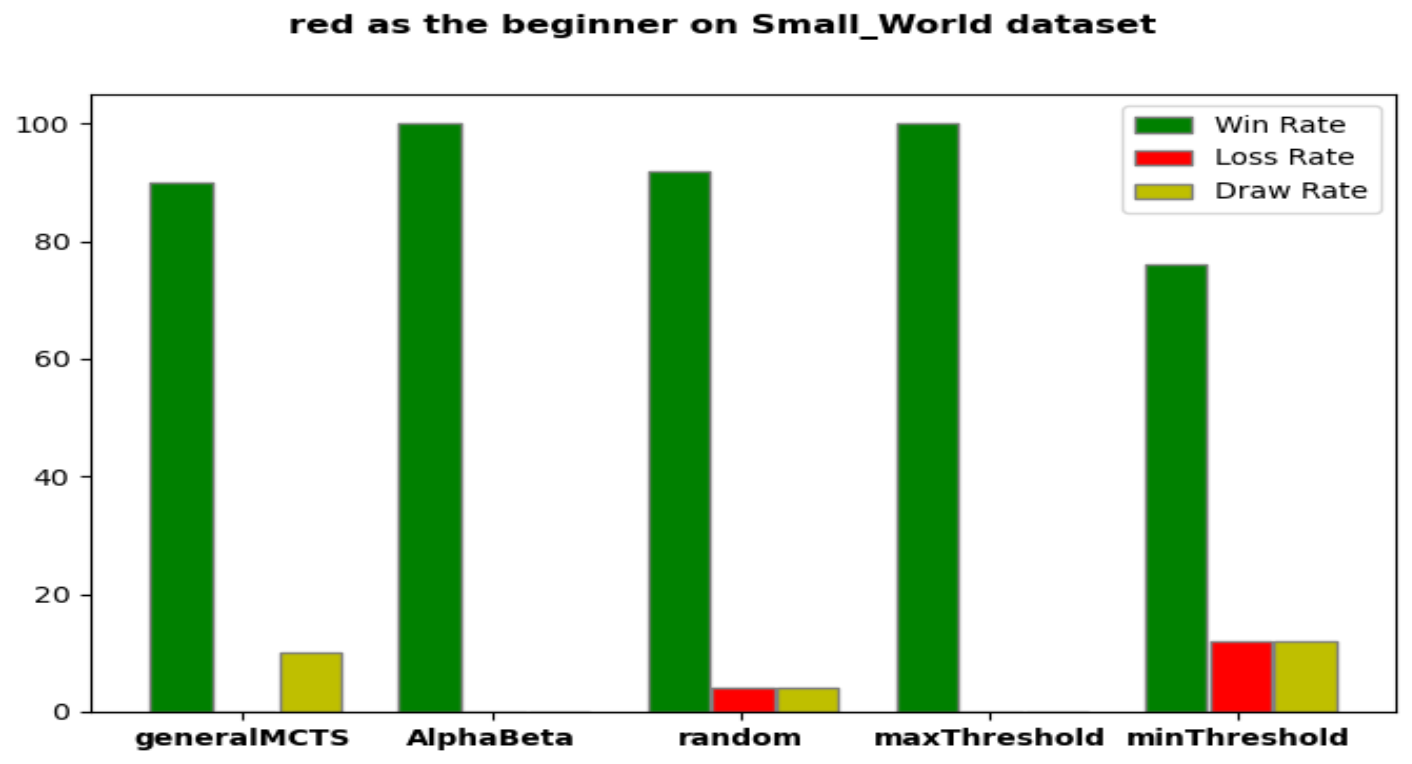

(b) Red player starts the game

**Fig. 5.** The results of the proposed strategy against other strategies on small-world graphs;

Fig. 5-a and 5-b demonstrate the results of different runs against different strategies on small-world graphs. Fig. 5-a shows the results when the black player starts the game, and Fig. 5-b shows the results of experiments when the red player starts. In both diagrams black strategy significantly outperforms against the different strategies of the red player. Playing against minimax algorithm with alpha beta pruning leads to the considerable wins over different runs whether black starts the game or red starts, this is due to the limitation in tree length that causes an incomplete search through the search space.

The minimax algorithm's output is even worse than random moves even though it takes lots of time to search through the tree to choose the best node. When the red player chooses max threshold, he defeats in all the game, which means max threshold is a bad choice for this problem; this is due to the limit on the number of tokens that causes choosing the node with the max threshold to be a wrong choice because a few nodes will be activated even if many neighbors got influenced through the diffusion process, but since they receive one token a few of them might be activated. On the other hand, the min threshold strategy, despite its simplicity, performs even better than general MCTS; this indicates that nodes with minimum threshold are a good choice because players' tokens will be saved and more nodes can be activated.

Table 2 indicates the amount of randomness on the results. Each column of the Table 2 shows the results of 20 times running the game on the same graph where black and red player choose ϵ-greedy MCTS and general MCTS strategy, respectively. It can be seen that the results are not deterministic, and there exists randomness in the results. For example, for the first sample, after 20 times running the game, two times red player wins, 17 times black player wins, and one time they are equal. However, the results are biased to a specific outcome (red player wins or black player wins or equal); but some randomness exists in the results.

**Table 3** Randomness of Scale-Free dataset on five graphs

| **Randomness** | **Graph 1** | **Graph 2** | **Graph 3** | **Graph 4** | **Graph 5** |
|---|---|---|---|---|---|
| Red Wins (%) | 60 | 70 | 25 | 5 | 30 |
| Black wins (%) | 20 | 15 | 65 | 80 | 25 |
| Draw (%) | 20 | 15 | 10 | 15 | 45 |

Table 3 shows the amount of randomness of the results for scale-free dataset. Accordingly, it can be seen that the results are not entirely definite, and there exists some randomness in the results. It can be concluded that, on average, 65% of the results have been biased to a specific result.

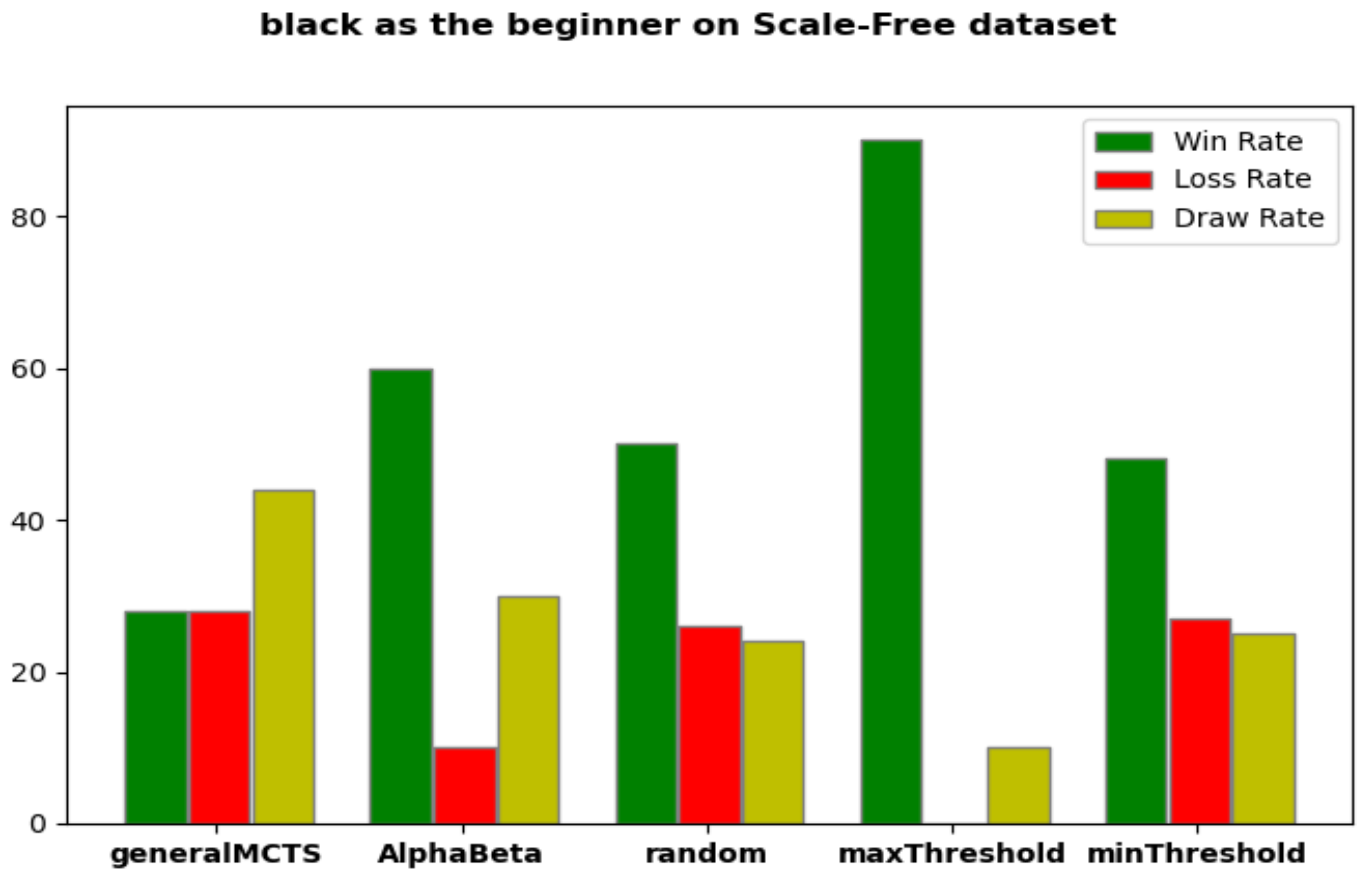

(a) Black player starts the game

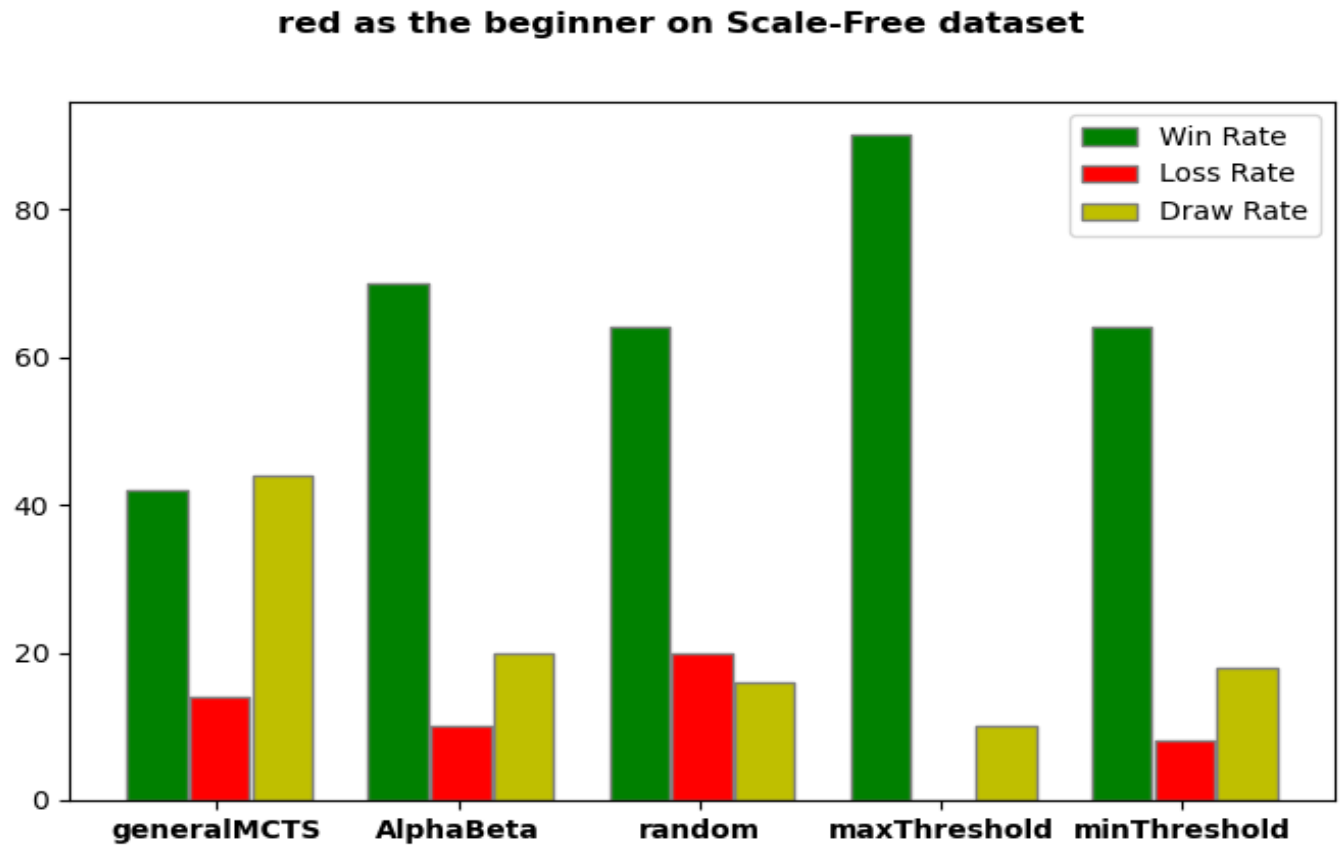

(b) Red player starts the game

**Fig. 6.** The results of the proposed strategy against other strategies on scale-free graphs;

The results on Scale-free datasets have been reported in Fig. 6-a and 6-b. As shown, these results are not as good as the small-world dataset results; the average nodes' degree in scale-free graphs are higher than the average nodes' degree of small-world graphs; therefore, players' tokens run out after a few numbers of nodes selection and player. It causes $\epsilon - greedy\ MCTS$ strategy cannot express its strength against the other strategies, while a low average degree in small-world graphs causes both players to select more nodes; therefore, the strategies can represent their strengths or weaknesses. Moreover, by comparing the above diagrams, one can realize that the results are significantly better when the red player (the opponent) starts the game. The reason is that by starting the game by the red player, it is so probable that the black player ends the game because the game is turn-base, but since the players do not donate the same amounts of tokens in each turn so it cannot be claimed that if one starts the game, the other one always ends it, but its more probable. On the other hand, because the nodes can change their state as much as the game goes on, the last move usually determines the winner of the game, so the player who ends the game has an ace up his sleeve.

**Table 4** Randomness of Random-Graph dataset on five graphs

| **Randomness** | **Graph 1** | **Graph 2** | **Graph 3** | **Graph 4** | **Graph 5** |
|---|---|---|---|---|---|
| Red Wins (%) | 60 | 60 | 45 | 10 | 15 |
| Black wins (%) | 15 | 15 | 15 | 70 | 60 |
| Draw (%) | 25 | 25 | 40 | 20 | 25 |

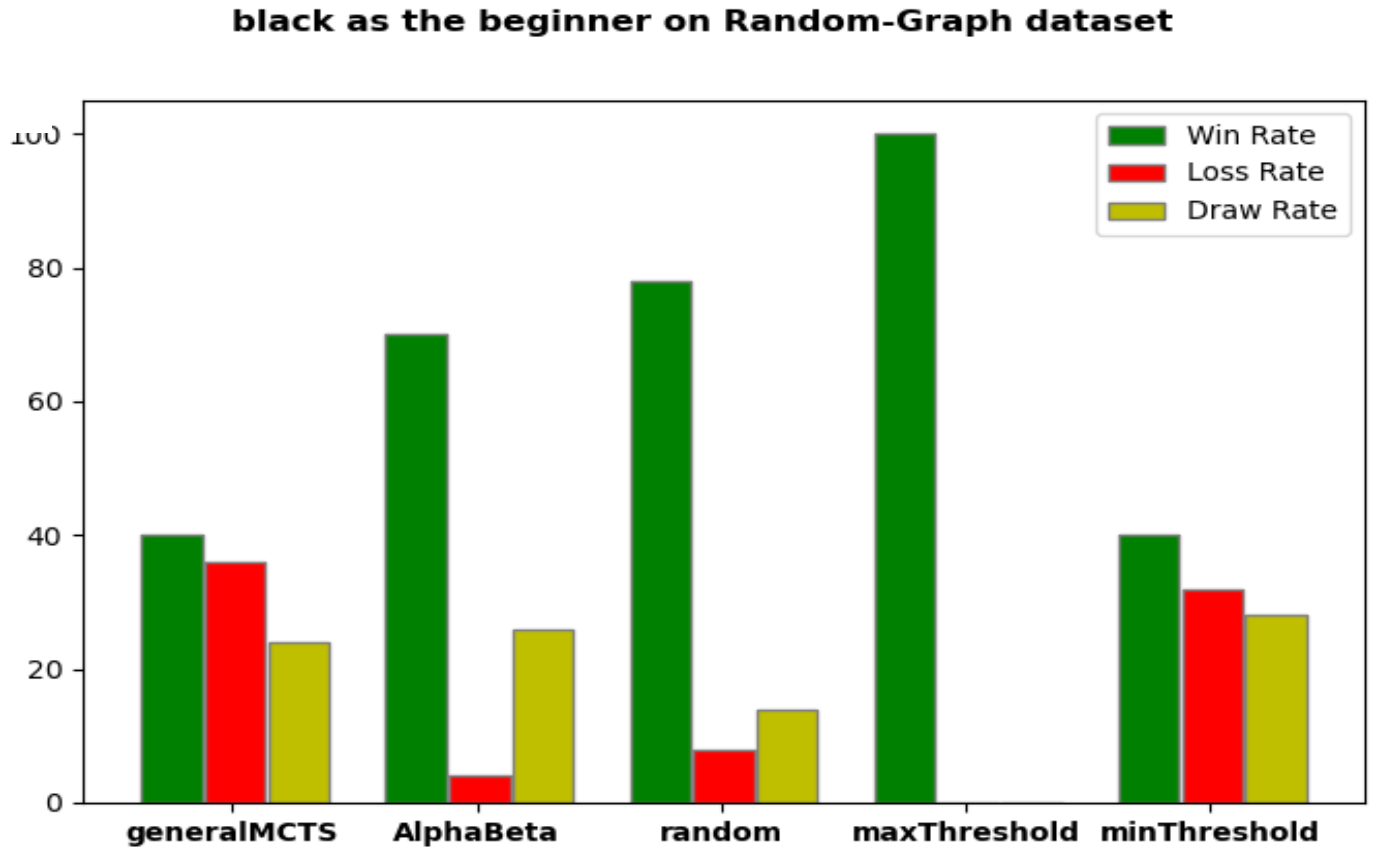

(a) Black player starts the game

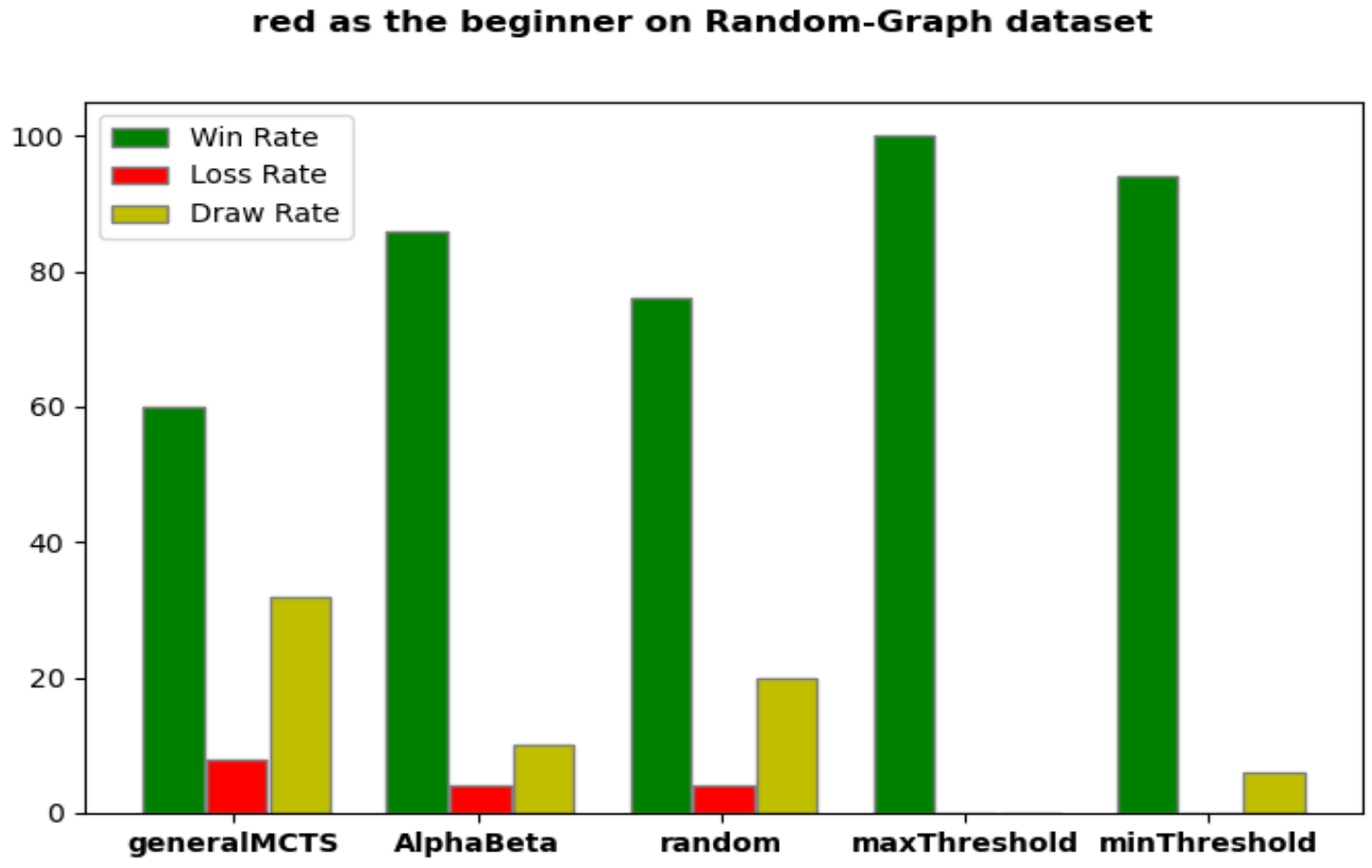

(b) Red player starts the game

**Fig. 7.** The results of the proposed strategy against other strategies on random graphs;

Fig. 7-a and 7-b demonstrate the results of the experiments on random graphs. Similar to the previous charts, the results are better when the red player starts the game. As mentioned, when the red player starts the game, black is most likely to end the game.

By investigating the Fig. 7-a and 7-b, it can be concluded that the win rate in random graphs is lower than small worlds graphs; the reason is related to the degree of nodes; since the graph's nodes degree are close to the average degree of the graph, which is equal to $(N-1) \times p$, each node is connected to approximately $p$ percent of the graph's nodes. This degree is higher than the nodes' degree in the small-worlds graphs, making the red and black players have limited choices, so the strategies cannot show their strengths or weaknesses. In addition, the high degree of nodes has led to high randomness in the final results on these graphs. By analyzing the Table 4 It can be concluded that the certainty in the results in random graphs is less than the small-worlds datasets and close to the scale-free ones, and on average, almost 60% of the time, the results are biased towards one of the states (red wins, black wins, draw). Due to the high degree of nodes, the $\epsilon - greedy\ MCTS$ strategy cannot show its superiority over other strategies and causes less certainty in the results. The results of competition against general MCTS and even minimum threshold algorithm demonstrate that when black starts the game, the $\epsilon - greedy\ MCTS$ is not capable of surpassing the opponents, however, minimax algorithm, random node and max threshold still fall behind of $\epsilon - greedy\ MCTS$. Table 4 demonstrates a considerable randomness in the results. As it can be derived, on average, 59% of the results have been biased to a specific result.

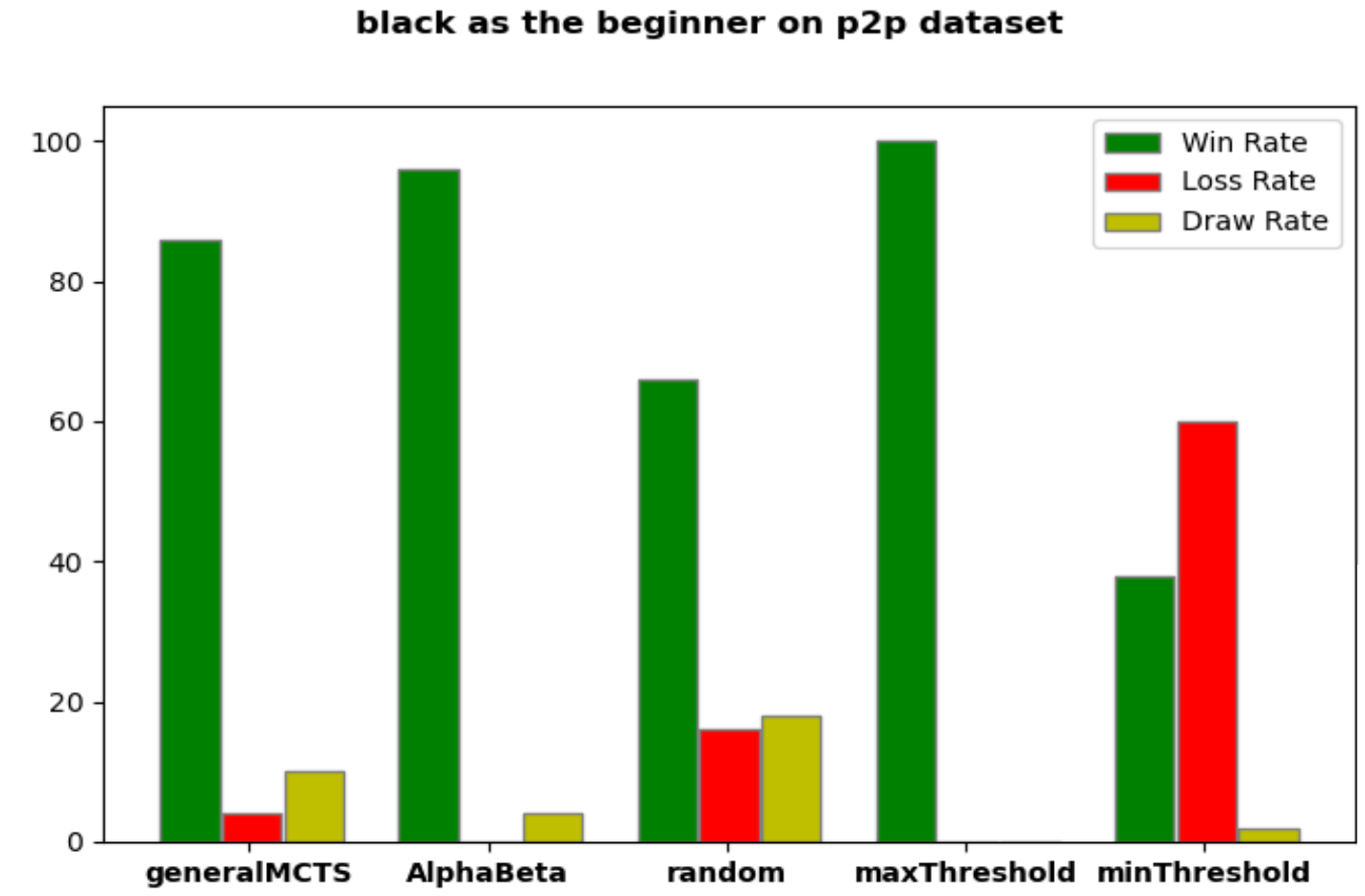

(a) Black player starts the game

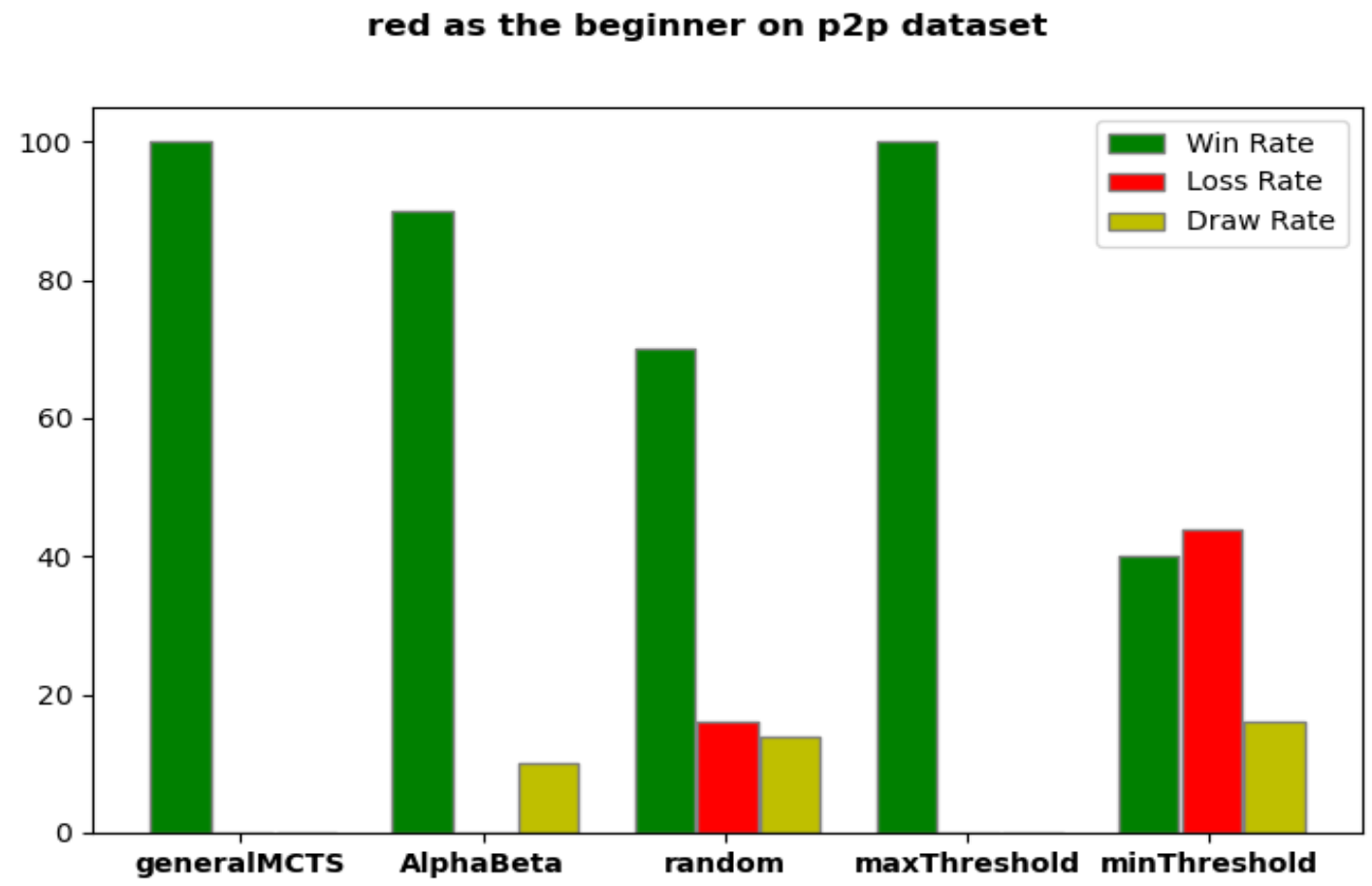

(b) Red player starts the game

**Fig. 8.** The results of the proposed strategy against other strategies on p2p.

Fig. 8-a and 8-b demonstrate the superiority of the $\epsilon - greedy\ MCTS$ strategy over the four strategies on p2p dataset. As mentioned earlier, the results are better when the red player is the starter than when black starts the game. A noteworthy point is the performance of the min threshold strategy, which performs better than $\epsilon - greedy\ MCTS$ in both diagrams and illustrates well the importance of nodes with low thresholds in the LBCIM framework. Despite the ingenuity of the $\epsilon - greedy\ MCTS$ algorithm, selecting the nodes with the lowest threshold in the proposed framework can have better or equal results in the selected cluster of the p2p database.

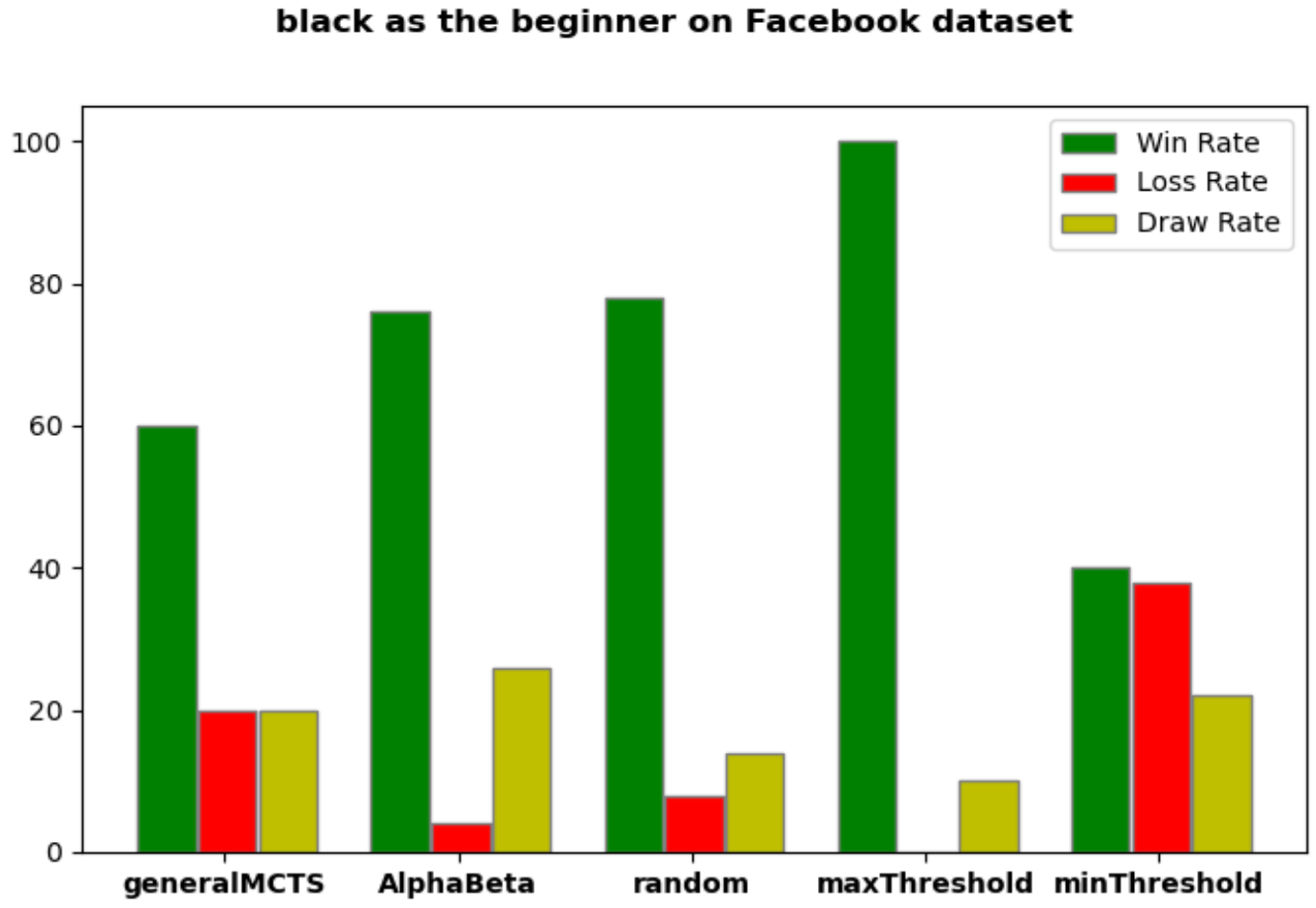

(a) Black player starts the game

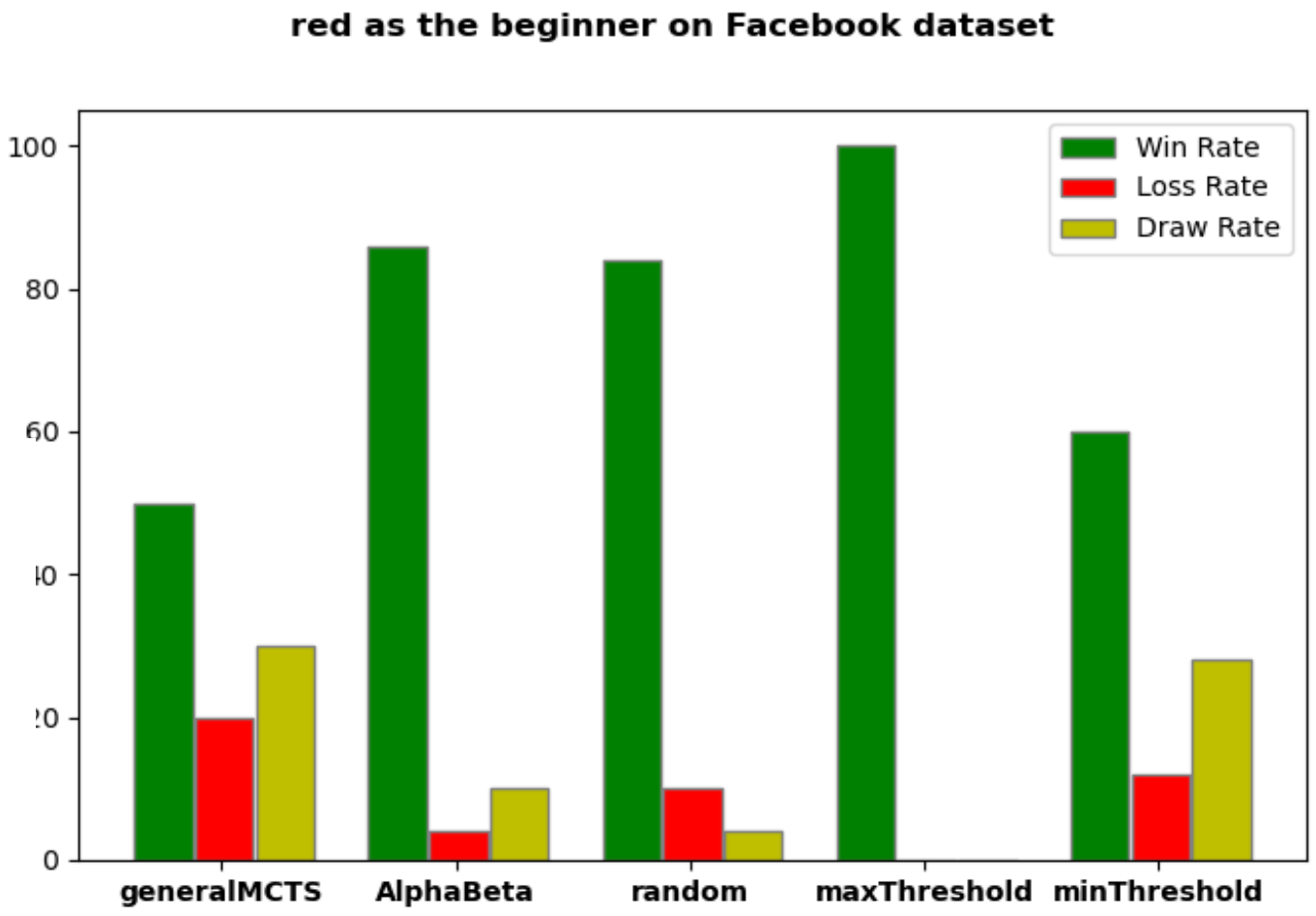

(b) Red player starts the game

**Fig. 9.** The results of the proposed strategy against other strategies on Facebook dataset;

Fig. 9-a and 9-b shows the results on Facebook database. It can be seen that the $\epsilon - greedy\ MCTS$ strategy performs better than the mentioned strategies. Contrary to expectations, the results against the MCTS strategy are better when the black player starts the game than when the red player starts. The reason is that although the game is played in turns, it cannot be stated with certainty that by starting a game by one party, the opponent always ends the game because the parties give a different number of tokens to the nodes. So, the party that awards the last tokens ends the game, not necessarily the rival party, but the rival party is more likely to end the game.

Randomness tables are not provided for these two datasets because the reported results are the results of 100 tests on one of the clusters of Facebook and the p2p network; therefore, the results also indicate the amount of randomness.

### 5.3 Discussion: How many tokens should be donated to the selected node?

In (Ishay et al. 2018), the game is implemented to denote one token to the selected node in each turn. In addition to imposing a lot of execution time, this manner performs unfavorably because the players invest their tokens on the selected node without any confidence that they can activate it in the next rounds of the game. Moreover, maybe the opponent in the next round activates this node. For this purpose, three different game formations are implemented and experimented on synthetic datasets.

1) The red player donates one token in each turn, while the black player donates some tokens equal to the selected node capacity to get fired. The capacity of node $u$ is defined as follow:

$$Capacity_u = \theta_u - (T_u^b + T_u^r) \; for \; u \in V \tag{14}$$

where $\theta_u$ is defined as the threshold of node $u$ and $T_u^b$ and $T_u^r$ are the black and red tokens of node $u$, respectively.

2) The red player donates one token to the selected node in each turn, while the black player can donate some amounts of tokens between one to the selected node's capacity based on the MCTS algorithm. Indeed, if the donated tokens of the black player to node $u$ are defined by $t$, it should be in followed range:

$$t \in [1\,, Capacity_u\,] \tag{15}$$

In this case, the children of one MCTS branch, in addition to including the eligible nodes of the graph, include all the combinations of tokens the black player can donate to these eligible nodes. In other words, if in a particular state of the game there exist $n$ eligible nodes for the black player and each of these nodes have a capacity of $m$ tokens on average, then that related branch has about $m \times n$ children.

3) Red player can decide to donate the amounts of tokens in the range $[1\,, Capacity_u]$ based on the MCTS algorithm, while black player donates its' tokens as much as the selected node capacity to get fired.

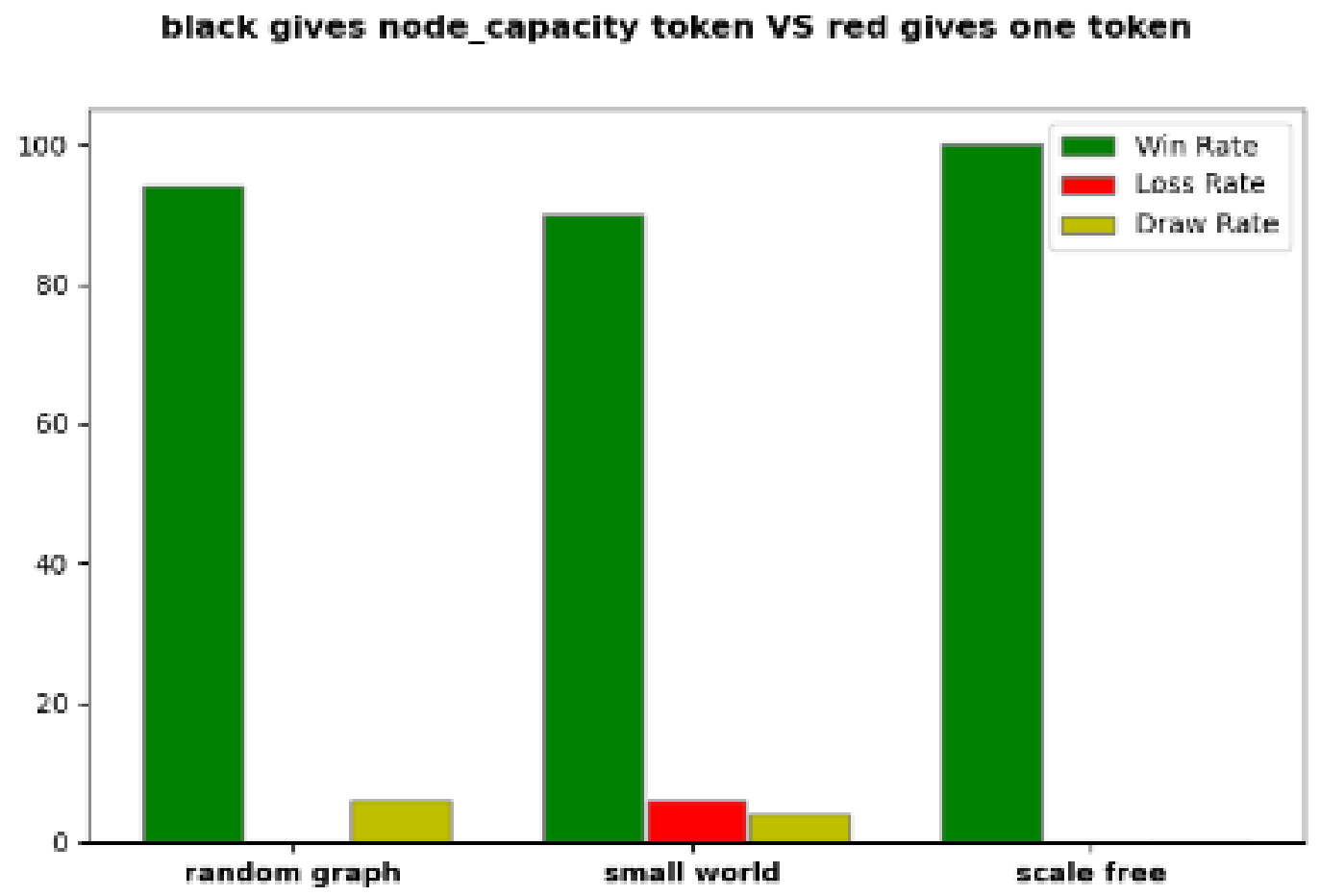

(a)

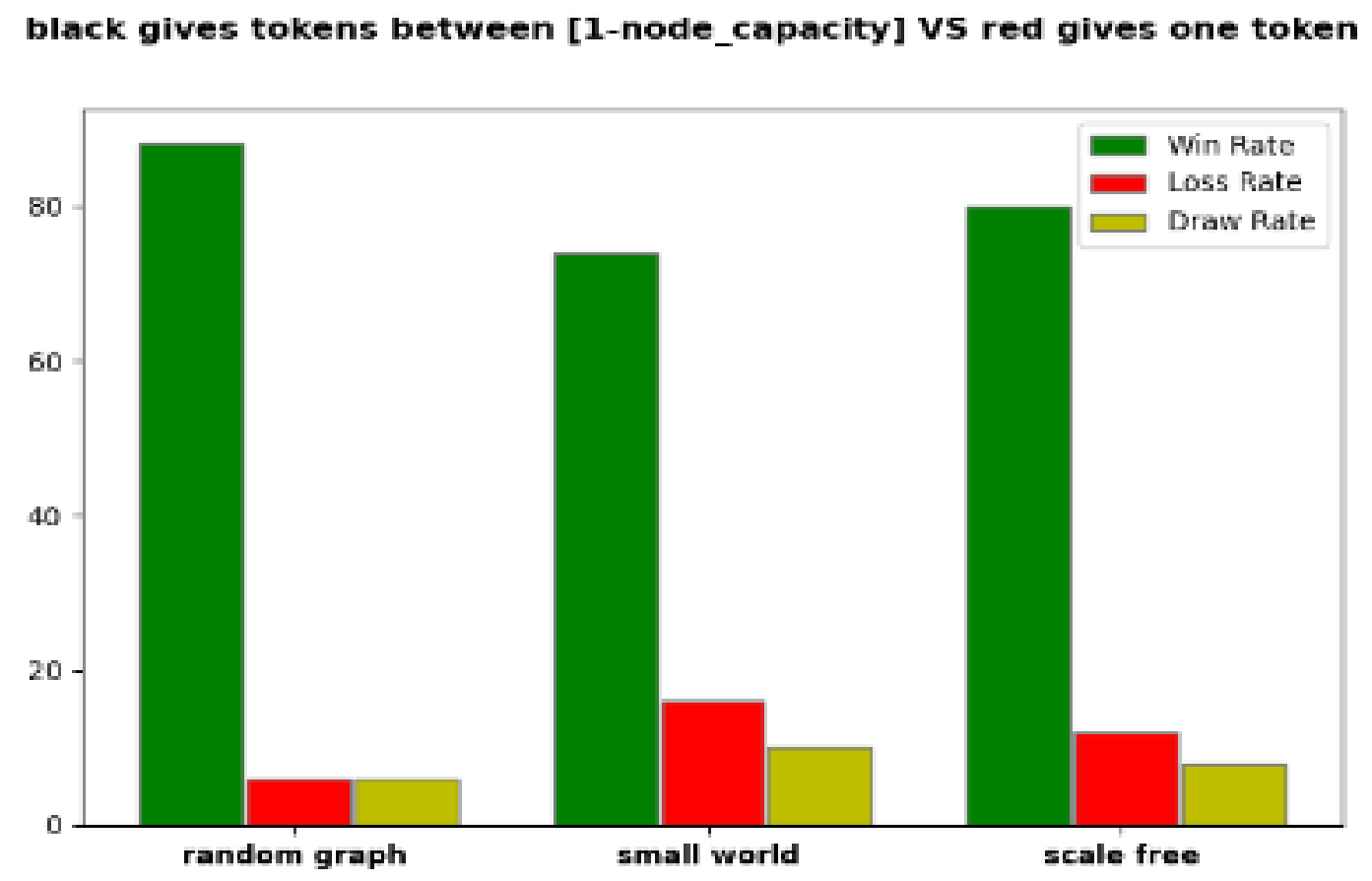

(b)

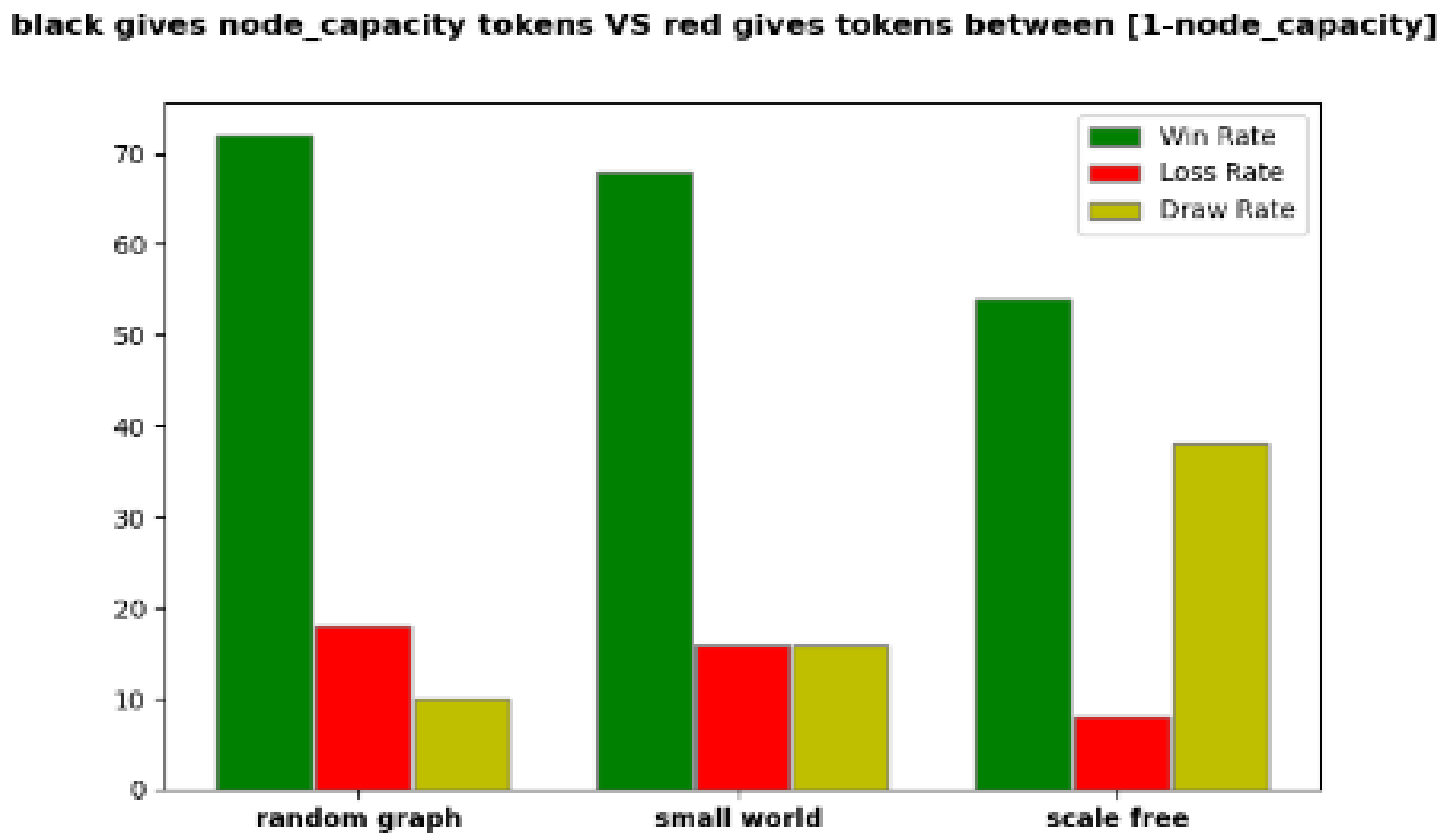

(c)

**Fig. 10.** The results of examining of donating different amounts of tokens to the selected node

Fig. 10-a shows the competition results between the two black and red parties in which black player fires the selected node at a time while red player donates only one token to the selected node. Both players use the Monte Carlo search algorithm to select nodes. As shown from Fig. 10-a, for all datasets, the black player wins the games by a considerable margin. It shows that if we give the nodes enough tokens to fire after selection, in addition to spending less time, better results will be obtained.

Fig. 10-b shows the competition results between the two parties, black and red, in which black chooses the optimal number of tokens for donating to the selected node with the help of the MCTS algorithm, and red only denotes one token at a time. Both players use the Monte Carlo search algorithm to select nodes. In this case, also, it can be seen that the value determined by the MCTS algorithm has a better result than when only one token is given to the node.

To better compare the two strategies, the Fig. 10-c examines when the black player fires the node after selecting, and the red player uses the MCTS algorithm to determine the optimal number of tokens for donating, which is between 1 to the capacity of the selected node. As shown from Fig. 10-c, the black player has performed better in the competition. the black strategy wins about 70% of the time against random-graphs and small-world datasets, and about 55% of the time, against the scale-free datasets. By comparing the results of the three above diagrams, it can be concluded that firing the node after selecting shows better performance.

## 6 Conclusion

In this paper, the impact of individuals' loyalty in a competitive environment has been examined. Each individual in such an environment can choose a party, and if it receives better offers from the rival, then it changes its mind and choose another party. As long as the game goes on, individuals are allowed to alternate their selected party. By knowing these individuals' properties, one can come up with good selections and win over the rivals. On the other hand, being aware of network properties and taking advantage of them have a considerable impact on the final result and surpass the competition.

The author has tried to cover all these environmental aspects in a competitive-based framework incorporating the loyalty factor. The experimental results indicate that the party equipped with the knowledge of network attributes significantly outperforms in node selections rather than the blind one with no such knowledge. Also, nodes loyalty has a massive impact on the determination of the winner, i.e., based on the results, nodes with low thresholds are better options because their lack of loyalty might be harmless than the high threshold nodes. Furthermore, the order of starting the game is also determinative due to the nodes' behavior.

For future works, one can develop the loyalty concept with a more complicated model that incorporates more aspects of loyalty or investigate the impact of different amounts of loyalty per node or add other factors like time and trust to the network. The competitive environment with multiple parties (more than two) can also be examined.